\documentclass[prd,showpacs,amsmath]{revtex4}
\begin{document}

\newcommand{\re}{\mathop{\mathrm{Re}}}

\newcommand{\be}{\begin{equation}}
\newcommand{\ee}{\end{equation}}
\newcommand{\bea}{\begin{eqnarray}}
\newcommand{\eea}{\end{eqnarray}}

\title{Statefinders and observational measurement of superenergy}

\author{Mariusz P. D\c{a}browski}
\email{mpdabfz@wmf.univ.szczecin.pl}
\affiliation{\it Institute of Physics, University of Szczecin, Wielkopolska 15, 70-451 Szczecin, Poland}
\author{Janusz Garecki}
\email{garecki@wmf.univ.szczecin.pl}
\affiliation{\it Institute of Mathematics, University of Szczecin, Wielkopolska 15, 70-451 Szczecin, Poland}

\date{\today}

\input epsf

\begin{abstract}
The superenergy of the universe is a tensorial quantity and it is a general relativistic
analogue of the Appell's energy of acceleration in classical mechanics. We propose the measurement of this quantity by the observational
parameters such as the Hubble parameter, the deceleration
parameter, the jerk and the snap (kerk) known as statefinders. We
show that the superenergy of gravity requires only the Hubble and
deceleration parameter to be measured, while the superenergy of matter requires also the
measurement of the higher-order characteristics of expansion: the jerk and the snap.
In such a way, the superenergy becomes another parameter characterizing the evolution of the universe.
\end{abstract}

\maketitle

\section{Introduction}

It is widely known that there exists a problem of the energy-momentum
of gravitational field in general relativity. Since
the gravitational field may locally vanish, then one is
always able to find a frame in which the energy-momentum of the gravitational
field vanishes, while it does not necessarily vanish in the other frames. The physical quantities which describe gravitational field in such a coordinate-dependent way are called the gravitational field pseudotensors, or, if the matter energy-momentum tensors are added, the gravitational field complexes. The choice of a gravitational field pseudotensor
is not unique so that various definitions of the
pseudotensors have been proposed \cite{pseudot}.
However, the arbitrariness of the choice of pseudotensors inspired many authors \cite{superenergy,supertensors} to define quantities which describe the energy-momentum
content of the gravitational field in a tensorial way. The two types of quantities have been suggested: the gravitational superenergy tensors \cite{superenergy} and the gravitational  super$^{(k)}$-energy tensors \cite{supertensors}. In particular, the canonical superenergy  tensors have successfully been calculated for the plane, the plane-fronted and cylindrical gravitational waves as well as for Friedmann, Schwarzschild, Kerr and G\"odel spacetimes
\cite{superenergy}. It was shown, for example, that a pure gravitational wave
with non-vanishing Riemann tensor components $R_{iklm}\not= 0$, possesses and carries positive-definite superenergy and that there exists a relation between the positivity of
superenergy and causality violation in G\"odel spacetimes. This means that the gravitational superenergy tensors are very useful quantities in studying the properties of gravity.

The definition of the superenergy tensor $S_a^{~b}(P)$, calculated at some spacetime point $P$, which can be applied to an arbitrary gravitational as well as a matter field is \cite{superenergy}
\begin{equation}
\label{Sabtet}
S_{(a)}^{~~~(b)}(y) = S_a^{~b}(P) \equiv \displaystyle\lim_{\Omega\to
P}{\int\limits_{\Omega}\biggl[T_{(a)}^{~~(b)}(y) -
T_{(a)}^{~~~(b)}(P)\biggr]d\Omega\over
1/2\int\limits_{\Omega}\sigma(P;y) d\Omega},
\end{equation}
where
\begin{eqnarray}
T_{(a)}^{~~~(b)}(y) &\equiv &
T_i^{~k}(y)e^i_{(a)}(y)e^{(b)}_k(y),\nonumber\\
\label{tetrad}
T_{(a)}^{~~~(b)}(P) &\equiv & T_i^{~k}(P)e^i_{(a)}(P)e^{(b)}_k(P) = T_a^{~b}(P)
\end{eqnarray}
are the tetrad components of a tensor or a
pseudotensor field $T_i^{~k}(y)$ which describe an energy-momentum, $y$
is the set of normal coordinates {\bf NC(P)} at a given point
{\bf P}, $\sigma(P,y)$ is the world-function, $e^i_{(a)}(y),~~e^{(b)}_k(y)$ denote an orthonormal tetrad field and its dual, respectively, $e^i_{(a)}(P) =
\delta^i_a,~~e^{(a)}_k(P) = \delta^a_k$, $e^i_{(a)}(y)e^{(b)}_i(y) =
\delta^b_a$, and they are paralell propagated along
geodesics through {\bf P}, and $\Omega$ is a spacetime volume element around {\bf P} (usually taken as the ball of radius $r$ in the limit $r \to 0$).

Following (\ref{Sabtet}), the canonical superenergy tensors of gravitation $_g S_i^{~k}$ and of matter $~_m S_i^{~k}$, which where explicitly introduced in a series of papers
listed in Ref. \cite{superenergy}, have the form
\begin{eqnarray}
_g S_i^{~k}(x;v^l) &=&  {2\alpha\over 9}\bigl(2v^nv^d -
g^{nd}\bigr)\bigl[4R^k_{~(lm)n}{}{R_i^{~lm}}_d\nonumber\\
&-& 2\delta_i^k{}R^{l(ma)}_{~~~~~n} R_{lmad} + 2\delta_i^k{} R^l_n
{}R_{ld}- 3 R^k_n{}R_{id} + 2 R^k_{~(i\vert m\vert n)} R^m_d\bigr],
\end{eqnarray}
\begin{equation}
_m S_i^{~k}(x;v^l)= h^{lm}{T_i^{~k}}_{;lm},
\end{equation}
where $x$ are local coordinates, $v^n$ is a four-velocity vector of an observer
{\bf O}; $h^{lm} \equiv 2v^lv^m - g^{lm}$ is an auxiliary
positive-definite Riemannian metric, $T_i^k$ are the components
of an energy-momentum tensor of matter, $R^i_{~klm}$ is the
Riemann curvature tensor, a semicolon denotes a covariant derivative and
a comma denotes an ordinary partial derivative.

Bearing in mind the differential nature of the definition (\ref{Sabtet}), it is interesting to note that the superenergy densities defined as $_g\epsilon = _g S_i^{~k}v^i v_k
~~_m\epsilon = _m S_i^{~k}v^i v_k$, calculated for an observer {\bf O} whose
four-velocity is $v^i$, correspond exactly to the Appell's {\it energy
of acceleration} $\frac{1}{2} m{\vec a}{\vec a}$ in non-relativistic mechanics. This energy of acceleration was first proposed over a century ago by Appell \cite{appell}.

In Ref. \cite{one} we studied the transformational
properties of both geometrical and physical quantities under conformal transformations of the metric. We noticed that in some conformal frames the quantities under studies are
of much simpler form than in original frames. Here we want to benefit from this and discuss  the superenergy tensors which can be conformally transformed to simpler forms. Besides, we propose a direct observational measurement of the local superenergy tensors by the application of some standard and non-standard cosmological parameters. In order to do so, we first  calculate conformal transformation rules for canonical superenergy tensors
of gravity ${_g S}_i^{~k}$ and matter ${_mS}_i^{~k}$.

In general, the conformal transformation rules for the superenergy
tensors are rather complicated. However, they vastly simplify when the transformed
metric is conformally flat and this is for example the case of standard Friedmann universes.

Our paper is organized as follows. We give the transformation rules
for superenergy tensors in Section \ref{section2}. In Section \ref{section3} we apply these rules to conformally flat Friedmann universes.
In Section \ref{section4} we express the non-zero components
of the superenergy tensors $_g S_i^{~k},~~_m
S_i^{~k}$ for a flat Friedmann universe {\it by
observational quantities} such as the Hubble parameter $H$, the
deceleration parameter $q$, the jerk $j$, and the snap (kerk) $k$.
We consider this as a crucial proposal of our paper since it gives
a regular mesurement of superenergy of the universe.
We will confine ourselves only to flat Friedmann universes and
we use geometrical units ($G=c=1$) and the signature $(+,-,-,-)$ throughout the paper.

\section{The transformation rules for the canonical superenergy
tensors}
\label{section2}

The conformal transformation defined as
\begin{equation}
\label{conf_trafo}
{\hat g}_{ik} = \Omega^2(x) g_{ik}(x),~~\Omega(x)>0,
\end{equation}
as applied to the canonical superenergy tensor of matter $_m S_i^{~k}$
written down in terms of metric $g_{ik}$, gives the superenergy tensor written down in terms of a conformal metric  ${\hat g}_{ik}$ in the form \cite{one}
\begin{eqnarray}
\label{Smat0}
_m {\hat S}_i^{~k}(x;{\hat v}^t)&=& {\hat h}^{lm}\bigl({\hat T}_i^{~k}\bigr)_{;{\hat l}{\hat m}}
 = {\hat h}^{lm}\bigl(\Omega^{-2}{}T_i^{~k}\bigr)_{;{\hat l}{\hat m}}\nonumber\\
&+&{\hat h}^{lm}\bigl(_c T_i^{~k}\bigr)_{;{\hat l}{\hat
m}} \equiv _m{\check S}_i^{~k}(x;{\hat v}^t) + _c{\hat S}_i^{~k}(x;{\hat
v}^t),
\end{eqnarray}
where
\begin{equation}
{\hat T}_i^{~k}(x) = \Omega^{-2}(x) T_i^{~k}(x) + _c T_i^{~k}(x) =
\Omega^2 {\tilde T}_i^{~k}(x) + _c T_i^{~k}(x), \hspace{0.3cm} {\tilde T}_i^{~k}(x) = \Omega^{-4}(x) T_i^{~k}(x)~,
\end{equation}
and the tensor ${\hat T}_i^{~k}$ is the total energy-momentum
tensor of matter in the new conformal frame ${\hat g}_{ik}(x)$,
$T_i^{~k}(x)$ are the components of the energy-momentum tensor of
matter in the old conformal frame, $g_{ik}$, and $_c T_i^{~k}(x)$ are the
components of an extra energy-momentum tensor of matter which is
created by the conformal transformation (\ref{conf_trafo}). It is worth mentioning that here
we took the Einstein equations in the new frame ${\hat g}_{ik}$ in the form
${\hat G}_i^{~k} = 8\pi {\hat T}_i^{~k}$ and in the initial
gauge  $g_{ik}$ in the form $G_i^{~k} = 8\pi T_i^{~k}$. Such a most natural procedure leads to the creation of matter by the conformal transformation (\ref{conf_trafo}).

The analytic form of the tensor $_c T_i^{~k}(x)$ is
\begin{equation}
\label{Tcreated}
_c T_i^{~k}(x) =
{1\over 8\pi}\bigl[{2\over\Omega}\bigl(\Omega^{-1}\bigr)_{;id}{}g^{kd}
+{\delta_i^k\over\Omega^3}\bigl(3\Omega_{;de}{}g^{de}
-{(\Omega^2)_{;ad}{}g^{ad}\over 2\Omega}\bigr)\bigr],
\end{equation}
and this tensor (\ref{Tcreated}) vanishes in the old frame,
$g_{ik}(x)$.
After some calculations we get
\begin{eqnarray}
\label{Smat}
_m{\check S}_i^{~k}(x;{\hat v}^t) &=& {\hat
h}^{lm}\bigl(\Omega^{-2} T_i^{~k}\bigr)_{;{\hat l}{\hat m}} =
\Omega^{-4}{} _m S_i^{~k}(x;v^t)+ \Omega^{-4}h^{lm}\bigl[\Omega^2{\Omega^{(-2)}}_{;lm}{}T_i^{~k}\nonumber\\
&+&\bigl(P^k_{~tm}{}T_i^{~t}\bigr)_{,l}-\bigl(P^t_{~mi}{}T_t^{~k}\bigr)_{,l}-2(ln\Omega)_{,l}P^k_{~tm}{}T_i^{~t}
-4(ln\Omega)_{,m}{}{T_i^{~k}}_{;l}\bigr]\nonumber\\
&+&\Omega^{(-5)}h^{lm}\bigl\{2(ln\Omega)_{,l} D^t_{~mi}{}T_t^{~k}
+\Gamma^k_{~lp}\bigl(D^p_{~tm}{} T_i^{~t}-
D^t_{~mi}{}T_t^{~p}\bigr)\nonumber\\
&-& \Gamma^p_{~lm}\bigl(D^k_{~tp}{}T_i^{~t}-
D^t_{~ip}{}T_t^{~k}\bigr) -
\Gamma^p_{~li}\bigl(D^k_{~mt}{}T_p^{~t}-
D^t_{~mp}{}T_t^{~k}\bigr)\nonumber\\
&-& D^p_{~lm}\bigl[{T_i^{~k}}_{;p}-2(ln\Omega)_{,p}{}
T_i^{~k}\bigr] +
D^k_{~lp}\bigl[{T_i^{~p}}_{;m}-2(ln\Omega)_{,m}{}T_i^{~p}\bigr]\nonumber\\
&-&D^p_{~li}\bigl[{T_p^{~k}}_{;m}-2(ln\Omega)_{,m}{}T_p^{~k}\bigr]\bigr\}\nonumber\\
&+& \Omega^{(-6)}h^{lm}\bigl[D^k_{~lp}\bigl(D^p_{~mt}{}T_i^{~t} -
D^t_{~mi}{}T_t^{~p}\bigr)-D^p_{~lm}\bigl(D^k_{~tp}{}T_i^{~t}-
D^t_{~ip}{}T_t^{~k}\bigr)\nonumber\\
&-& D^p_{~li}\bigl(D^k_{~mt}{} T_p^{~t} - D^t_{~ip}{}
T_i^{~k}\bigr)\bigr],
\end{eqnarray}
where
\begin{equation}
D^a_{~bc} = \Omega P^a_{~bc} = \delta_b^a \Omega_{,c} +
\delta^a_c \Omega_{,b} - g_{bc} g^{ad} \Omega_{,d} ~,
\end{equation}
and
\begin{eqnarray}
\label{Sc}
_c{\hat S}_i^{~k} (x;{\hat v}^l)& =& {\hat h}^{lm} {_c
T_i^{~k}}_{;{\hat l}{\hat m}} = \Omega^{-2}h^{lm}\bigl[ {_c
T_i^{~k}}_{;ml} + \bigl(P^k_{~tm}{}_c T_i^{~k}\bigr)_{,l} -
\bigl(P^t_{~mi}{}_c T_i^{~k}\bigr)_{,l}\nonumber\\
&+&\Gamma^k_{~lp}\bigl(P^p_{~tm}{}_c T_i^{~t} -P^t{~mi}{}_c
T_t^{~p}\bigr)- \Gamma^p_{~lm}\bigl(P^k_{~pt}{}_c T_i^{~t} -
P^t_{~ip}{}_c T_t^{~k}\bigr)\nonumber\\
&-& \Gamma^p_{~li}\bigl(P^k_{~mt}{}_c T_p^{~t}- P^t_{~mp}{}_c
T_t^{~k}\bigr)+ P^k_{~lp}{_c T_i^{~p}}_{;m}\nonumber\\
&+& P^k_{~lp}\bigl(P^p_{~mt}{}_c T_i^{~t} - P^t_{~mi}{}_c
T_t^{~p}\bigr) - P^p_{~lm}{_c T_i^{~k}}_{;p}\nonumber\\
&-& P^p_{~lm}\bigl(P^k_{~tp}{}_c T_i^{~t} - P^t_{~ip}{}_c
T_t^{~k}\bigr) - P^p_{~li}{_c T_p^{~k}}_{;m}\nonumber\\
&-& P^p_{~li}\bigl(P^k_{~mt}{} _c T_p^{~t} - P^t_{~mp}{}_c
T_t^{~k}\bigr)\bigr].
\end{eqnarray}
Here $_m S_i^{~k}$ is the canonical superenergy tensor of matter in
the old frame $g_{ik}(x)$, and the tensor $_c{\hat S}_i^{~k}$
is the superenergy tensor of the matter which is created by the
conformal transformation (\ref{conf_trafo}). The latter tensor
vanishes in the old frame, $g_{ik}(x)$.
In order to obtain the transformation rule for the canonical superenergy
of matter (under the conformal transformation (\ref{conf_trafo})) one has to insert  (\ref{Smat}) and (\ref{Sc}) into the right-hand-side of equations (\ref{Smat0}).

Now, following \cite{one}, the transformation rule for the gravitational canonical superenergy tensor $_g S_a^{~b}$ reads as
\begin{eqnarray}
\label{Sgrav}
_g{\hat S}_a^{~b}(x;{\hat v}^l)&=& \Omega^{-4} _g S_a^{~b}(x;v^l)
+ {2\alpha\over 9}h^{lm}\bigl\{\bigl[\bigl(g^{k[b}\Omega^{i]}_{~~l}
-\delta^{[b}_l{}\Omega^{i]k}\bigr)\nonumber\\
&+&\bigl(g^{i[b}{}\Omega^{k]}_{~~l}-\delta^{[b}_l{}\Omega^{k]i}\bigr)\bigr]g_{[a[k}{}\Omega_{i]m]}\nonumber\\
&-& {\delta_a^b\over 2}\bigl(g^{k[i}{}\Omega^{j]}_{~~l}
-\delta^{[i}_l{}\Omega^{j]k}\bigr)\bigl(g_{[i[k}{}\Omega_{j]m]}
+g_{[i[j}{}\Omega_{k]m]}\bigr)\bigr\}\nonumber\\
&+&  {2\alpha\over 9}\Omega^{-1}h^{lm}(\Omega^{-1})_{;mc}\bigl(g^{bc}{}\Omega_{al}-
\delta^b_{(a}{}\Omega^c_{~l)} +g_{al}\Omega^{bc} -
\delta^c_{(a}{}\Omega^b_{~l)}\bigr)\nonumber\\
&+&{4\alpha\over 9}\Omega^{-2}h^{lm}\bigl[
2R^{b(ik)}_{~~~~~l}{}g_{[a[k} \Omega_{i]m]} +
\bigl(g^{k[b}{}\Omega^{i]}_{~~l}-\delta^{[b}_l{}\Omega^{i]k}\bigr)R_{a(ik)m} \nonumber\\
&-& \delta_a^b{}R^{i(jk)}_{~~~~~l}{}g_{[i[k}\Omega_{j]m]}-{\delta_a^b\over
2}\bigl(g^{k[i}{}\Omega^{j]}_{~~l}-\delta^{[i}_{l}{}\Omega^{j]k}\bigr)R_{i(kj)m}\nonumber\\
&+&4\delta_a^b(\Omega^{(-1)})_{;lg}{}(\Omega^{-1})_{;mc}{}
g^{gc}-6(\Omega^{-1})_{;al}{}(\Omega^{-1})_{;m}^{~~b}\nonumber\\
&+& {1\over 4}\bigl(R^b_{~m}{}\Omega_{al} -
R^g_{~m}{}\delta^b_{(a}{}\Omega_{l)g} +
g_{al}{}R^g_{~m}{}\Omega^b_{~g}-
R_{m(a}{}\Omega^b_{~l)}\bigr)\bigr]\nonumber\\
&+&{4\over 3}\alpha\Omega^{-3} h^{lm}\bigl[{4\over 3}\delta^b_a
(\Omega^{(-1)})_{;cm}{}R^c_{~l}-(\Omega^{(-1)})_{;m}^{~~b}{}R_{al}\nonumber\\
&-&(\Omega^{(-1)})_{;al}{}R^b_{~m} +{1\over
3}(\Omega^{(-1)})_{;mc}\bigl({R^b_{~a}}^c_{~l}+
{R^b_{~l}}^c_{~a}\bigr)\bigr]\nonumber\\
&+&{\alpha\over 18}\Omega^{-4}
h^{lm}(\Omega^2)_{;rs}{}g^{rs}\bigl(\delta^b_{[a}{}\Omega_{l]m} +
g_{l[m}{}\Omega^b_{~a]}\bigr)\nonumber\\
&+& {2\over 3}\alpha\Omega^{-5}
h^{lm}(\Omega^2)_{;rs}{}g^{rs}\bigl[(\Omega^{(-1)})_{;al}{}\delta^b_m
+ g_{al}(\Omega^{(-1)})_{;m}^{~~b}-{4\over 3}\delta^b_a
(\Omega^{(-1)})_{;ml}\bigr]\nonumber\\
&+&{\alpha\over 3}\Omega^{-6}
h^{lm}(\Omega^2)_{;rs}{}g^{rs}\bigl(R_{al} \delta^b_m +
g_{al}{}R^b_{~m} -{4\over 3}\delta^b_a{}R_{lm} -{1\over
3}\delta^g_m{}R^b_{~lga}\bigr)\nonumber\\
&+&{\alpha\over 9}\Omega^{-8} h^{lm}\bigl(\Omega^2\bigr)_{;rs}
g^{rs}\bigl(\Omega^2\bigr)_{;uv}{}g^{uv}\bigl(\delta_a^b{}g_{lm}
-{3\over 2}\delta^b_m{}g_{al}\bigr).
\end{eqnarray}
Here $_g S_a^{~b}$  are the components of the canonical
superenergy tensor of gravitation in the old frame, $g_{ik}$, and $\alpha = {1/(16\pi)}$.
As one can see, the transformational rules for the superenergy tensors of matter and
gravitation are are quite complicated. However, if the initial metric, $g_{ik}(x)$, is Minkowskian, i.e., if we confine ourselves to a conformally flat spacetime with metric ${\hat g}_{ik}= \Omega^2(x)\eta_{ik}$, then the above transformational formulas
simplify significantly because, in the initial frame, $g_{ik}(x)$,
one has
\begin{equation}
\label{mink}
g_{ik}= \eta_{ik},~~T_i^{~k} = 0, ~~_m S_i^{~k} = 0, ~~_g S_a^{~b}
   = 0,~ R^i_{~klm}=0,~~ R_{ik}=0,~~R=0,~~_{;i} = _{,i}.
\end{equation}
In consequence, we have in this case that
\begin{equation}
_m{\hat S}_i^{~k}(x;{\hat v}^t) = _c {\hat S}_i^{~k}(x;{\hat
v}^t),
\end{equation}
because $_m{\check S}_i^{~k}(x;{\hat v}^t) \equiv 0$, and $_c {\hat S}_i^{~k}(x;{\hat v}^t)$
is substantially simplified in comparison to (\ref{Sc}) due to the conditions
(\ref{mink}).

\section{The case of a flat Friedmann universe}
\label{section3}

Friedmann universes are conformally flat and, in a special case of its their flat geometry, we have
\begin{equation}
\label{FRWm}
ds^2 = a^2(\tau)\bigl(d\tau^2 - dx^2 - dy^2 -dz^2\bigr),
\end{equation}
which means that the conformal factor is just the scale factor \cite{Gar2,Jap}) $\Omega= \Omega(\tau) = a(\tau) > 0$, and $g_{ik} = \eta_{ik}$ in (\ref{conf_trafo}).
It is important here that in (\ref{FRWm}) the conformal time $\tau$
has been introduced.

We would like to emphasize that one can obtain a flat, and also
curved Friedman universes with all their energetic and superenergetic content
from the flat Minkowski spacetime by a suitable conformal
transformation (see e.g. \cite{one} for details). An analogous statement is true for any other conformally flat spacetime so that we do not need any quantum fluctuations of the Minkowski vacuum to create a Friedmann or any other
conformally flat spacetime - the idea which is commonly used in quantum cosmology
(see e.g. \cite{kieferQG}). Amazingly, a classical conformal
transformation of the initial Minkowski metric is sufficient to do the job.

The components of the superenergy tensors $_g {\hat S}_i^{~k}(x;v^t)$
and $_m {\hat S}_i^{~k}(x;v^t)$ for the Friedmann metric (\ref{FRWm}) can be calculated by using the formulas (\ref{Smat0}),(\ref{Smat}),(\ref{Sc}),(\ref{Sgrav}), simplified by (\ref{mink}), and give
\begin{eqnarray}
\label{gS00}
_g {\hat S}_0^{~0}& =& {\alpha\over 9a^8}\bigl(204 {a^{\prime}}^2 -
396a{a^{\prime}}^2 a^{\prime\prime} + 188a^2
{a^{\prime\prime}}^2\bigr),\\
\label{gS11}
_g {\hat S}_1^{~1}&=& _g {\hat S}_2^{~2} = _g {\hat S}_3^{~3} = {\alpha\over
3a^8}\bigl(10{a^{\prime}}^4 - 22a {a^{\prime}}^2 a^{\prime\prime}
+{20\over 3}a^2 {a^{\prime\prime}}^2\bigr),\\
\label{mS00}
_m {\hat S}_0^{~0} &=& {12\alpha \over
a^6}\bigl({a^{\prime\prime}}^2 + a^{\prime} a^{\prime\prime\prime}
- {16{a^{\prime}}^2 a^{\prime\prime}\over a} +
{22{a^{\prime}}^4\over a^2}\bigr),\\
\label{mS11}
_m {\hat S}_1^{~1}& =&_m {\hat S}_2^{~2}   = _m {\hat S}_3^{~3} =
{4\alpha \over a^5}\bigl(a^{\prime\prime\prime\prime}
{11a^{\prime} a^{\prime\prime\prime}\over a}- {4
{a^{\prime\prime}}^2\over a} + {40{a^{\prime}}^2
a^{\prime\prime}\over a^2} - {22{a^{\prime}}^4\over a^3}\bigr),
\end{eqnarray}
where $a^{\prime} \equiv da/ d\tau, ~~a^{\prime\prime} \equiv d^2a/
d\tau^2$,~~$a^{\prime\prime\prime} \equiv d^3a/ d\tau^3$,~~
$a^{\prime\prime\prime\prime} \equiv d^4a/ d\tau^4$,
For a zero pressure dust-filled flat Friedmann universe one gets
\begin{equation}
\Omega(\tau) = a(\tau) = {A^3\over 9} \tau^2, ~~0<\tau<\infty,
\end{equation}
where
\begin{equation}
A \equiv \bigl(6\pi\rho a^3\bigr)^{1/3} = const>0~,
\end{equation}
and we have from (\ref{gS00})-(\ref{mS11})
\begin{eqnarray}
\label{gS00f}
_g {\hat S}_0^{~0}&=& {618192\alpha\over A^{12} \tau^{12}}>0,~~_g
{\hat S}_1^{~1} = _g {\hat S}_2^{~2} = _g {\hat S}_3^{~3}=
{23328\alpha\over A^{12}\tau^{12}}>0,\\
\label{gS11f}
_g {\hat S} &:=& _g {\hat S}_0^{~0} + _g {\hat S}_1^{~1} + _g {\hat
S}_2^{~2} + _g {\hat S}_3^{~3} = {688176\alpha\over A^{12}
\tau^{12}}>0,\\
\label{mS00f}
_m {\hat S}_0^{~0}&=& {8975448\alpha\over
A^{12}\tau^{12}}>0,~~_m {\hat S}_1^{~1} = _m {\hat S}_2^{~2} = _m
{\hat S}_3^{~3} = - {1259712 \alpha\over\kappa A^{12}
\tau^{12}}<0,\\
\label{mS11f}
_m {\hat S}&:=& _m {\hat S}_0^{~0} + _m {\hat S}_1^{~1} + _m {\hat
S}_2^{~2} + _m {\hat S}_3^{~3} = {1417176 \alpha\over\kappa A^{12}
\tau^{12}}>0.
\end{eqnarray}


\section{Observational measurement of superenergy}
\label{section4}

Bearing in mind the final output for superenergy tensors given by (\ref{gS00})-(\ref{mS11}),
we notice that they can be expressed in terms of some standard and non-standard cosmological parameters. The well-known characteristics of the universe expansion are
the Hubble parameter $H$, and the deceleration parameter $q$:
\bea
\label{hubb+dec}
H = \frac{\dot{a}}{a}~, \hspace{1.cm} q  =  - \frac{1}{H^2} \frac{\ddot{a}}{a} = - \frac{\ddot{a}a}{\dot{a}^2}~,
\eea
while the new characteristics are the jerk $j$ \cite{jerk}, and the snap (kerk) \cite{snap}
\bea
\label{jerk+snap}
j = \frac{1}{H^3} \frac{\dddot{a}}{a} =
\frac{\dddot{a}a^2}{\dot{a}^3}~,\hspace{1.cm} k = -\frac{1}{H^4} \frac{\ddddot{a}}{a} =
-\frac{\ddddot{a}a^3}{\dot{a}^4}~.
\eea
A general form of these parameters can be expressed as \cite{Mar}
\bea
\label{dergen}
x^{(i)} &=& (-1)^{i+1}\frac{1}{H^{i}} \frac{a^{(i)}}{a} = (-1)^{i+1}
\frac{a^{(i)} a^{i-1}}{\dot{a}^{i+1}}~,
\eea
with $i$ - the natural number, `(i)' means the ith derivative while `i' is just a power.

It is obvious that nowadays we can measure $H$ and $q$ \cite{supernovae}. However,
the observational determination of the value of jerk by
using type Ia supernovae sample is more difficult. Despite that it has been performed
\cite{riess2004} and the claim is that $j_0 > 0$. Similar
investigations may perhaps also be possible for higher-order
characteristics such as kerk/snap, lerk etc. Some hints about
that are given in \cite{snap}.

Now, our objective is to express the non-vanishing components of the
canonical superenergy tensors (\ref{gS00})-(\ref{mS11}) in terms of the
above observational parameters (\ref{hubb+dec}) and (\ref{jerk+snap}).
Since the standard definition of these parameters uses the derivatives of the scale factor
with respect to the {\it cosmic time} $t$ instead of the derivatives with respect to
the conformal time $\tau$, we will use the conformal time definition formula
\begin{equation}
\label{taut}
{\dot \tau}\equiv {d\tau\over dt}= {1\over a(t)},
\end{equation}
to express all the derivatives with respect to the cosmic time $t$
by the derivatives with respect to the conformal time $\tau$ and
rewrite the observational parameters $H,~q,~j,~k$ in terms of the
derivatives with respect to the conformal time $\tau$. By doing
so, one can express the non-zero components of the superenergy tensors (\ref{gS00})-(\ref{mS11}) in terms
of the observational parameters $H,~q,~j,~k$ as
\begin{eqnarray}
\label{Sgh1}
_g{\hat S}_0^{~0} &=& {4\alpha H^4\over 9}\bigl(47q^2+
5q-1\bigr),\\
\label{Sgh2}
_g{\hat S}_1^{~1}&=& _g{\hat S}_2^{~2} = _g{\hat S}_3^{~3} =
{2\alpha H^4\over 9}\bigl(10q^2+13q -8\bigr),\\
\label{Smh1}
_m{\hat S}_0^{~0}&=& {12 \alpha H^4}\bigl(q^2+10q
+j+8\bigr),\\
\label{Smh2}
_m{\hat S}_1^{~1} &=&_m{\hat S}_2^{~2} = _m{\hat S}_3^{~3}
={4 \alpha H^4}\bigl(q -4j-k+4\bigr).
\end{eqnarray}
Theoretically, for a dust-filled flat Friedman universe one has $q=1/2,~~j=1,~~k=7/2$. So, in this case both superenergy densities $_g{\hat S}_0^{~0}$ and $_m{\hat S}_0^{~0}$
for a comoving observer {\bf O} {\it are positive-definite}.
One can see from the formulas (\ref{Sgh1})-(\ref{Smh2}) that the
measurement of the superenergy of gravity requires only the measurement of the Hubble parameter $H$ and the deceleration parameter $q$, while the measurement of the superenergy of matter
requires also the measurement of the jerk, $j$, and the snap (kerk), $k$.

We claim that our formulas (\ref{Sgh1})-(\ref{Smh2}) are of great physical
importance because they {\it allow a direct measurement} of
the components of the canonical superenergy tensors for
Friedmann universes from large-scale cosmological observations.

Using (\ref{taut}) and taking into account the observational values of the
$H_0 = 71 km/sMpc$, $q_0 = -0.55$, $j_0 >0$, we then obtain the values of the superenergy near the point of measurement
on Earth at the present moment of the evolution of the universe. Also, by using
the rules of the time-evolution of these parameters we may express the superenergy
throughout the whole evolution of the universe. This means that both superenergy tensors
are observational quantities and {\it can be added to a set of the standard
parameters characterizing the evolution of the universe}.

\section{Conclusion}

We have calculated the conformal transformation rules for
the canonical superenergy tensors, gravitation
and matter. The general rules we
obtained are pretty complicated, but they vastly simplify
when the spacetime under study is conformally flat. This happens,
for example, for the Friedman universes.

We have applied our general formulas to the flat Friedman
universes and obtained relatively simple expressions for the
components of the canonical superenergy tensors of matter and
gravitation. Finally, we have expressed these
components in terms of the observational parameters: the Hubble parameter, the
deceleration parameter, the jerk and the snap (kerk).

According to current observations, our Universe is likely to be the flat Friedman universe.
Then the formulas (\ref{Sgh1})-(\ref{Smh2}) we obtained, allow {\it a direct measurement} of the superenergy densities of the Universe from large-scale cosmological observations.

\acknowledgments
We acknowledge the support of the Polish Ministry of
Science and Higher Education grant No N N202 1912 34 (years 2008-10).


\begin{thebibliography}{99}

\bibitem{pseudot} A. Einstein, Preuss. Ak. Wiss. Sitzungsber., 270 (1918); L.D. Landau and E.M. Lifschitz {\it The Classical Theory of Fields} (Oxford: Pergamon, 1975); C. M\o ller {\it The Theory of Relativity} (Oxford: Clarendon, 1972); A. Papapetrou {\it Proc. R. Ir. Acad.} {\bf A 52}, (1948) 11; P.G. Bergmann and R. Thomson {\it Phys. Rev.} {\bf 89}, 401 (1953); S. Weinberg {\it Gravitation and Cosmology} (Wiley, New York, 1972); D. Bak, D. Cangemi and R. Jackiw, Phys. Rev. D{\bf 49}, 5173 (1994).


\bibitem{supertensors} J.M.M. Senovilla, Class. Quantum Grav. {\bf 17} (2000)
2799; J.M.M. Senovilla, Mod. Phys. Lett. A{\bf 15}, 159 (2000); P. Teyssandier, e-print gr--qc/9905080.

\bibitem{superenergy} F.A.E. Pirani, Phys. Rev. {\bf 105} (1957)
1089; J.Garecki, Rep. Math. Phys. {\bf 33}, 57 (1993);
Int. J. Theor. Phys. {\bf 35}, 2195 (1996); Rep.
Math. Phys. {\bf 40}, 485 (1997); Journ. Math. Phys. {\bf 40}, 4035
(1999); Rep. Math. Phys. {\bf 43}, 397 (1999);
Rep. Math. Phys. {\bf 44}, 95 (1999); Ann. Phys. (Berlin),
{\bf 11}, 441 (2002); Class. Quantum Grav. {\bf 22}, 4051 (2005);
Found. Phys. {\bf 37}, 341 (2007); M.P. D\c
abrowski and J. Garecki, Class. Quantum Grav. {\bf 16}, 1 (2002).

\bibitem{one} M.P. D\c abrowski, J. Garecki, D.B.
Blaschke, Ann. Phys. (Berlin) {\bf 18}, 13 (2009); M.P. D\c{a}browski and J. Garecki, arXiv: 0712.1358.

\bibitem{appell} P.E. Appell, Journ. f\"ur die reine und angewandte Mathematik {\bf 121}, 310 (1900).

\bibitem{Gar2} J. Garecki, Acta Phys.Pol. {\bf B 39}, 781
(2008).

\bibitem{Jap} M. Iihoshi et al., Prog.Theor.Phys. {\bf
118}, 475 (2007).

\bibitem{kieferQG} C. Kiefer, {\em Quantum Gravity} (Oxford University Press, 2007).

\bibitem{Mar} M.P. D\c abrowski, Phys. Lett. {\bf B 625}, 184
(2005).

\bibitem{jerk} E.R. Harrison, Nature {\bf 260}, 591 (1976);
               P.T. Landsberg, Nature {\bf 263}, 217 (1976);
               T. Chiba, Prog. Theor. Phys. {\bf 100} (1998), 1077;
               Yu. Shtanov and V. Sahni Class. Quantum Grav. {\bf
               19}, L101 (2002);
               U. Alam, V. Sahni, T.D. Saini, and A.A. Starobinsky, Mon. Not. R.
               Astron. Soc. {\bf 344}, 1057 (2003); V. Sahni, T.D.
               Saini, A.A. Starobinsky, and U. Alam JETP Lett.
               {\bf 77}, 201 (2003);
               M. Visser, Class. Quantum Grav. {\bf 21}, 2603 (2004).

\bibitem{snap} R.R. Caldwell and M. Kamionkowski, JCAP 0409 (2004), 009;
M.P. D\c{a}browski and T. Stachowiak, Annals of Physics (N.Y.) {\bf 321} (2006), 771.

\bibitem{supernovae} J.L. Tonry {\it et al.\/}, Astroph. J. {\bf 594}, 1
(2003); M. Tegmark {\it et al.\/}, Phys. Rev. D {\bf 69}, 103501
(2004); P. Astier {\it et al.\/}, Astron. Astrophys.
{\bf 447}, 31 (2006); A.G. Riess {\it et al.\/}, Astroph.J, {\bf 659}, 98, 2007;
M.W. Wood-Vasey {\it et al.\/}, Astroph. J., {\bf 666}, 694, 2007;
M. Kowalski et al. Astroph. J., {\bf 686}, 749 (2008).

\bibitem{riess2004} A.G. Riess {\it et al.\/}, Astroph. J. {\bf 607}, 665 (2004).


\end{thebibliography}
\end{document}